\documentclass[
 preprint,
 amsmath,amssymb,
 aps,
]{revtex4-2}

\usepackage{graphicx}
\usepackage{dcolumn}
\usepackage{bm}
\usepackage{hyperref}
\usepackage{subfigure}
\usepackage{xcolor}

\begin{document}


\title{Scaling and Condensation of Dry Active Matter Around Circular Obstacles}

\author{Felipe P. S. Júnior}
\email{felipe.junior@icen.ufpa.br}
\affiliation{%
Faculdade de F\'\i sica, Universidade Federal do Par\'a ICEN, Caixa Postal 479, 66075-110, Bel\'em, Par\'a, Brazil
}%

\author{F. Q. Potiguar}
\email{fqpotiguar@ufpa.br}
\affiliation{%
Faculdade de F\'\i sica, Universidade Federal do Par\'a ICEN, Caixa Postal 479, 66075-110, Bel\'em, Par\'a, Brazil
}%
 \author{Jorge L. C. Domingos}%
\affiliation{%
MMML Lab, Department of Physics, University of Latvia, Jelgavas 3, Riga, LV-1004, Latvia
}%

\author{W. P. Ferreira}%
\affiliation{%
 Departamento de F\'\i sica, Universidade Federal do Cear\'a, Caixa Postal 6030, 60455-760, Fortaleza, Cear\'a, Brazil.
}%




\date{\today}

\begin{abstract}
Active Brownian particles confined to rigid substrates are known to accumulate near rigid
boundaries and, under suitable conditions, undergo motility-induced phase separation (MIPS). A
particularly intriguing manifestation of this behavior is the formation of self-sustained vortices
around circular obstacles, which act as localized nucleation sites for particle aggregation. While
several dynamical properties of such vortices have been previously characterized, their behavior in
the thermodynamic limit remains largely unexplored. Here, we investigate how the mass and spatial
extent of a dry active-matter vortex scale with system size. Using numerical simulations of
repulsive active Brownian Particles interacting with a fixed circular obstacle, we measure the
vortex mass, mean radius, and maximum radius as functions of the global area fraction, obstacle
size, and system size. We find two distinct scaling regimes. At low densities, the vortex remains
localized and its characteristic properties saturate as the system size increases. Above a critical
density, however, the vortex mass grows extensively with the total number of particles, while its
spatial dimensions scale linearly with the system size, indicating the emergence of an obstacle-stabilized condensed state. Cluster-size distributions reveal that this transition occurs at densities
significantly below those associated with bulk MIPS, demonstrating that a localized geometric
heterogeneity can strongly modify the condensation scenario of active matter. To rationalize these
observations, we propose a minimal kinetic interpretation based on a balance between particle
capture from the dilute phase and particle escape driven by rotational diffusion. Our results show
that circular obstacles act as efficient nucleation centers capable of stabilizing macroscopic active
aggregates and provide a framework for understanding condensation phenomena in geometrically
confined active systems.
\end{abstract}

\maketitle


\section{Introduction\label{Intro}}

Active matter consists of self-driven units that continuously consume energy from their
surroundings and convert it into motion, thereby maintaining systems far from thermodynamic
equilibrium \cite{marchetti13,tevrugt24,volpe25}. Examples range from bacterial colonies and motile cells to artificial
microswimmers and active colloids \cite{marchetti13,berke08}. One of the most remarkable collective phenomena
observed in active systems is Motility-Induced Phase Separation (MIPS), in which purely repulsive
self-propelled particles spontaneously segregate into dense and dilute phases despite the absence of
attractive interactions. 
Over the last decade, MIPS has become one of the central paradigms
of active matter, motivating extensive theoretical, numerical, and experimental investigations \cite{fily12,redner13,buttinoni13,fily14,cates15,stenhammar15,wittmann16,liebchen17,digregorio18,caprini20,xiaqing20,su2021,chacon22,rojasvega23,caporusso23}.

The physical mechanism underlying MIPS originates from a positive feedback between particle
accumulation and motility reduction, which requires a sufficiently high density to develop \cite{cates15}.
Persistent particles slow down in crowded regions, which promotes further accumulation and
eventually leads to the emergence of macroscopic dense aggregates \cite{fily12,volpe25}. In systems of Active
Brownian Particles (ABPs), bulk phase separation is observed only above a critical density that
depends on activity, typically occurring at area fractions around $\varphi \simeq 0.5$ in the high-Péclet number regime \cite{fily12,buttinoni13,liebchen17,volpe25}. Below this threshold, homogeneous states remain stable and only
finite clusters are expected \cite{buttinoni13,liebchen17}. As a direct consequence of this mechanism, true bulk MIPS is
not expected at low area fractions: densities of the order of $\varphi \simeq 0.20$ lie well below the
theoretical instability threshold, and any clustering observed in this regime must instead be
attributed to finite-size effects, boundaries, or heterogeneous nucleation processes rather than
genuine bulk phase coexistence.

Closely related to phase separation is the behavior of active particles close to solid surfaces. The
presence of boundaries and geometric heterogeneities can profoundly modify this scenario \cite{pan20,bendor22}. Active
particles are known to accumulate near walls, corners, and obstacles because their persistent motion
increases the residence time close to confining surfaces \cite{caporusso23,sepulveda17,sepulveda18,turci21, cates15,fily14}. This phenomenon gives rise to
active wetting \cite{caporusso23,sepulveda17, sepulveda18,turci21}, boundary-induced clustering \cite{stenhammar15,caporusso23,sepulveda17, sepulveda18,turci21}, and rectification effects in
asymmetric environments \cite{bechinger16,makarchuk19,doostmo16}. In this context, obstacles may act as heterogeneous nucleation
centers capable of stabilizing dense active structures even when the global density lies below the
threshold required for bulk phase separation.

Among the structures induced by confinement, active vortices formed around circular obstacles
constitute a particularly interesting example. Active vortices have been reported both in wet
systems, where they emerge from collective hydrodynamic interactions and mesoscale turbulence
\cite{galajda07,mahmud09,wioland16,heidenreich16,james21,bricard15}, and in dry systems, where they arise from the persistent motion of self-propelled particles
interacting with confining boundaries and obstacles \cite{schimming25,tiwari24,potiguar14}. In particular, previous studies have
shown that dry active particles can spontaneously organize into rotating aggregates surrounding a
fixed circular obstacle \cite{schimming25,tiwari24}. The formation mechanism, rotational dynamics, correlations
between neighboring vortices, the control of  vortex rotation using asymmetric neighboring constraints have been investigated in detail \cite{mokhtari17,pan20,junior24,junior26}, revealing that obstacle
geometry plays a crucial role in determining vortex stability and collective behavior. Nevertheless, a
fundamental question remains unanswered: how does such a vortex behave as the system size
increases?

This question is closely related to the role played by a localized nucleation site in the
thermodynamic limit. Since a vortex is formed by particles captured from the surrounding dilute
phase, one may ask whether a finite obstacle can sustain a macroscopic aggregate when the
available space becomes arbitrarily large. Equivalently, does the vortex remain a localized structure
whose mass saturates with system size, or can it become an extensive condensate containing a finite
fraction of all particles in the system? More generally, how does the presence of a preferred
nucleation site alter the conventional picture of active phase separation? Addressing these questions
is important for understanding how geometric confinement competes with bulk phase separation
and whether heterogeneous nucleation can stabilize condensed active states below the density
threshold associated with bulk MIPS.

In this work, we address these questions by studying the scaling properties of dry active-matter
vortices formed around a circular obstacle. Using numerical simulations of Active Brownian
Particles with purely repulsive interactions, we analyze the dependence of the vortex mass, mean
radius, and maximum radius on the system size, particle density, and obstacle diameter. We show
that the system exhibits two qualitatively distinct regimes. At low densities, the vortex remains
localized and its characteristic properties saturate as the system size increases. Above a critical
density, however, the vortex becomes extensive, with its mass scaling linearly with the total number
of particles and its spatial extent growing proportionally to the system size.

These findings indicate that a localized geometric heterogeneity can stabilize condensed active
aggregates at densities substantially lower than those required for bulk MIPS. Our results therefore
provide insight into how confinement and heterogeneous nucleation modify the collective
organization of active matter and establish a connection between active vortices, obstacle-induced
condensation, and finite-size scaling in non-equilibrium systems.



In Section \ref{Model}, we describe our model and simulation details; in Sec. \ref{Results}, we show our results and conclude our paper in Sec. \ref{Conclusions}.

\section{Model Description\label{Model}}
We consider a collection of $N$ active particles of unit diameter $\sigma$ and self-propulsion speed $v_0$ in a square box of length $L$, with periodic boundaries. In the center of this box there is a fixed circular obstacle of diameter $D>\sigma$.

Two particles $i$ and $j$, located at ${\bf r}_i$ and ${\bf r}_j$ respectively, interact through a linear spring force law ${\bf F}_{ij}=\kappa(d_{ij}-r_{ij}){\hat{\bf r}}_{ij}$ for $d_{ij}>r_{ij}$, and ${\bf F}_{ij}=0$ otherwise, where $d_{ij}=\frac{1}{2}(\sigma_{i}+\sigma_{j})=\sigma$ and $r_{ij}=|{\bf r}_i-{\bf r}_j|$ is the distance between the two particles; $\kappa$ is the spring stiffness. For a particle interacting with the obstacle, we have $\kappa_{obs}>\kappa$, and $d_{ij}=\frac{1}{2}(\sigma+D)$.

A particle $i$ moves according to the overdamped Langevin equation:
\begin{equation}
    \label{eqn1}
    {\bf v}_i=\frac{d{\bf r}_i}{dt}=\mu{\bf F}_i+{\bf u}_i+{\bf A}_i(t),
\end{equation}
where ${\bf F}_i=\sum\limits_{j}{\bf F}_{ij}$ is the total force on particle $i$, $\mu$ is the mobility, ${\bf A}_i(t)$ is a Gaussian random variable with zero mean and correlation intensity  $\xi$; ${\bf u}_i=v_0[\sin\theta_i(t){\bf i}+\cos\theta_i(t){\bf j}]$ is the intrinsic velocity of the particle $i$, whose direction $\theta_i(t)$ follows the stochastic equation:
\begin{equation}
    \label{eqn2}
    \frac{d\theta_i}{dt}=\eta_i(t).
\end{equation}
The quantity $\eta(t)$ is also a Gaussian random variable with zero mean and correlation magnitude $\eta=2D_r$, where $D_r$ is the rotational diffusion coefficient. We consider the athermal version of the full model; therefore, we set $\xi=0$. Hence, we obtain results for an infinite Péclet number $\mathrm{Pe}=v_0\sigma/\mu k_BT$ \cite{digregorio18}. Both equations are integrated simultaneously with a second-order stochastic Runge-Kutta algorithm \cite{honeycutt92}.

The model parameters are given as $\mu=1$, $\kappa=50$, $\eta=0.001$. As indicated above, our unit length is one active particle diameter $\sigma=1$, and the time unit is set such that the  self-propulsion speed is unity ($v_0=1$). We consider four values of diameter of the obstacle, $D=10,20,30$ and $40$. This range of obstacle sizes covers the three reported vortex regimes \cite{pan20} with respect to the vortex motion, namely random state ($D\leq10$), transient state ($10<D\leq20$), and vortex state ($D>20$), for the density values considered. We consider various values of box length $L$, for fixed $D$, keeping the area fraction of the particles $\varphi=\frac{N\pi}{4(L^2-\pi D^2/4)}$ constant. The  area fraction is varied over the range $\varphi=[0.05,...,0.50]$. 
The number of particles used in our simulations ranges from $N=80$ to $N=64400$. We measured our observables over, typically, $10-30$ independent runs, depending on the overall computational cost; each run is for $3000$ time units, with $2000$ time units used to achieve the steady state.

Our main objective is to measure three distinct properties of the vortex, namely, its mean number of particles, or mass, $\overline{M}$, its mean radius, $\overline{R}$ and its mean maximum radius, $\overline{R}_{max}$. In order to do this, we first identify the particles that are connected to the obstacle. We consider two particles to be connected if their center-to-center distance is less than $\sigma$; for particle-obstacle connection, their center-to-center distance should be less than half the sum of their diameters, $(\sigma+D)/2$. We consider a cluster whenever at least two particles are connected. In turn, a cluster is part of the vortex if at least one of its particles is connected to the obstacle. Hence, if a cluster $C$, with mass $m(C)$, is attached to the obstacle, the total mass of the vortex $V$ (formed by all the attached clusters) is defined as
\begin{equation}
    \label{eqn3}
    \overline{M}=\left<\sum\limits_{C} m(C)\right>.
\end{equation}
The mean radius is calculated as
\begin{equation}
    \label{eqn4}
    \overline{R}=\left<\frac{1}{M}\sum\limits_{i\in V} |{\bf r}_i-{\bf R_{0}}|\right>,
\end{equation}
where the sum is taken over all the particles in the vortex $V$, $M$ is the mass of the vortex and ${\bf R_{0}}=\frac{L}{2}({\bf i}+{\bf j})$ is the center of the box, {\em i.e.}, the position of the obstacle and the angle brackets indicate averages over distinct realizations. Finally, the mean maximum radius, which is the distance of the farthest particle in the vortex to the obstacle, is calculated as
\begin{equation}
    \label{eqn5}
    \overline{R}_{max}=\left<\mathrm{max}|{\bf r}_i-{\bf R_{0}}|\right>,~~~~~~i\in V.
\end{equation}

\section{Results\label{Results}}
We will begin by presenting our results for the mean vortex mass, followed by our kinetic model; then, we show the Cluster Size Distribution data, and, finally, the data for the mean and maximum radii.

\subsection{Scaling of the vortex mass}
\begin{figure}[b]
\includegraphics[scale=1.0]{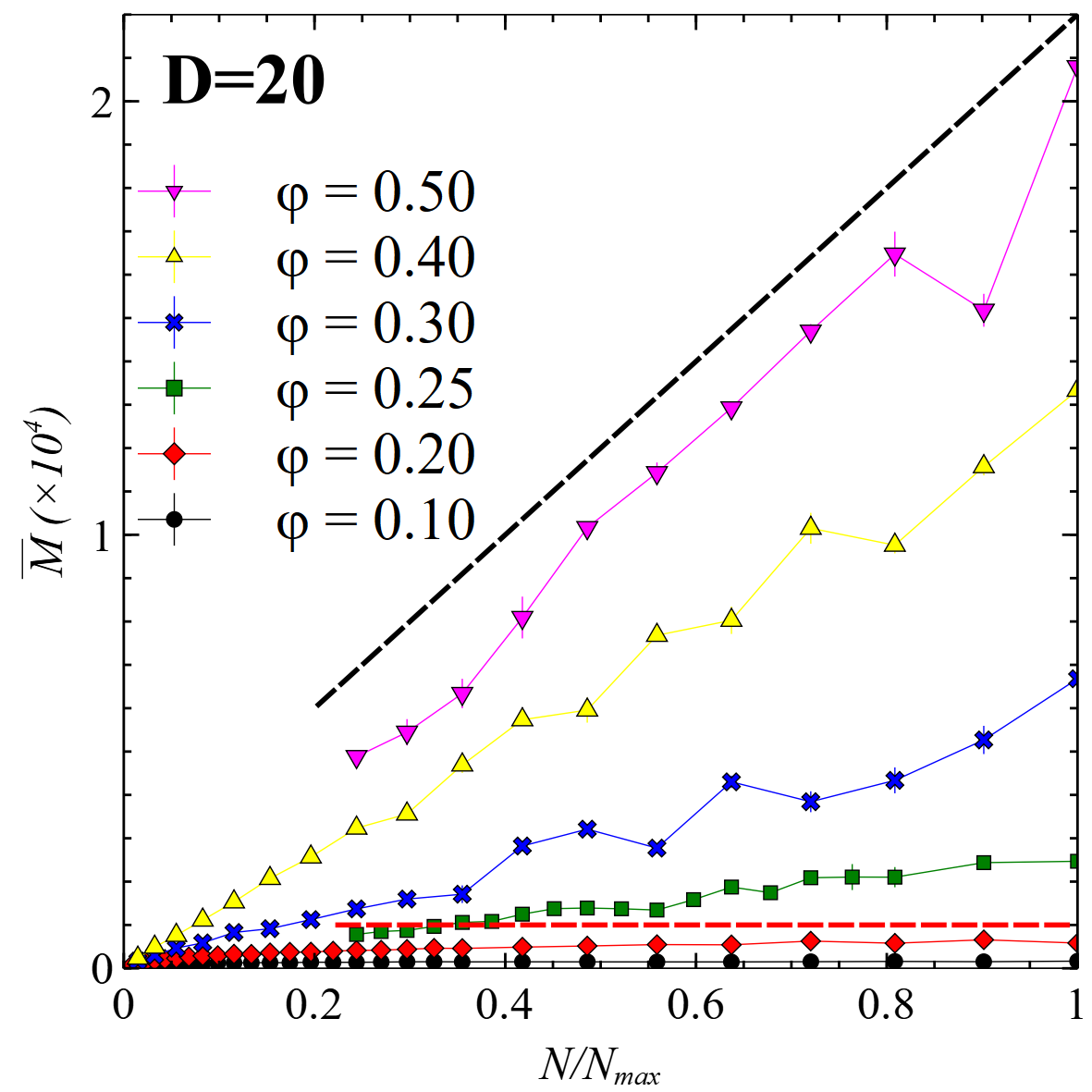}
\caption{Total vortex mass, $\overline{M}$, as a function of the total number of particles normalized by the number of particles at $L/D=10$ for different densities $\varphi$. The black and red dashed lines are $\overline{M}\sim(N/N_{max})$ and $\overline{M}=\mathrm{const}$, respectively.\label{fig1}}
\end{figure}

\begin{figure}[b]
\includegraphics[scale=1.0]{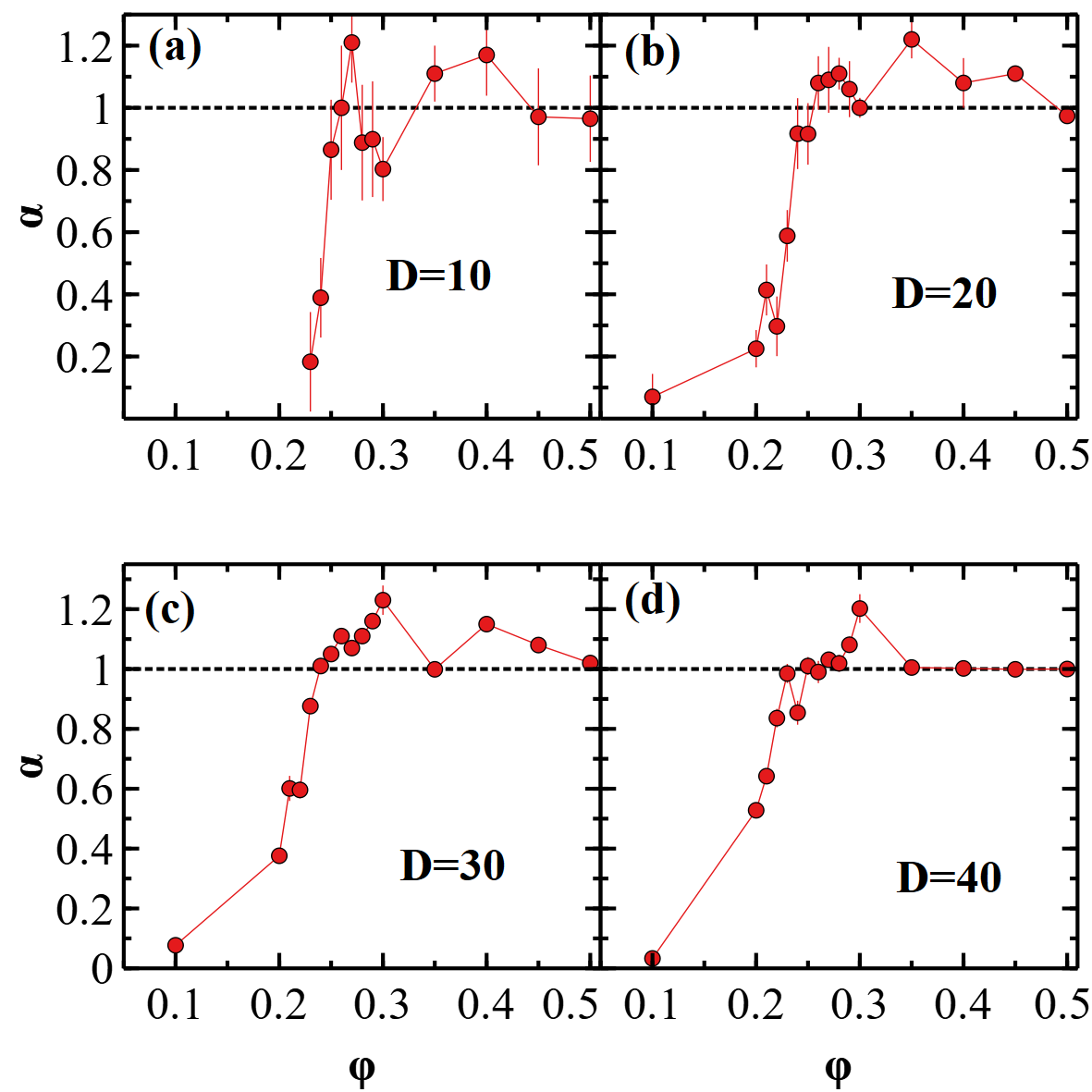}
\caption{Mass exponent $\alpha$, as a function of the density for different obstacle sizes: (a) $D=10$, (b) $D=20$, (c) $D=30$, and (d) $D=40$.\label{fig2}}
\end{figure}

Let us begin with the results for the vortex mass as a function of the total number of particles $N$, for $D=20$ and for different $\varphi$. In Fig. \ref{fig1}, we show $\overline{M}$ as a function of $N/N_{max}$, where $N_{max}$ is the number of particles when $L/D=10$, which is the largest ratio of system size to obstacle diameter we will consider. We show the data in this way for a better visualization. We see that as we increase the density,  $\overline{M}$  increases more rapidly with $N$;  the data indicate that it goes from nearly saturating, at large $N$, for $\varphi=0.10$ (compare the data with the red dashed line) up to an approximate linear dependence at $\varphi=0.50$ (compare to the black dashed line).
We observe similar behavior  for other obstacle sizes. Hence, we can assume that the relationship between them follows a power law 
\begin{equation}
\label{eqn6}
    \overline{M}\sim N^{\alpha(\varphi)}
\end{equation}
in which the exponent $\alpha$ depends on the density $\varphi$; the linear regime $\alpha=1$ corresponds to phase separation in a model of chiral swimmers \cite{liebchen17}, where a solid-like cluster coexists with fluid-like phase, see Fig. \ref{fig3}. We show the measured values of $\alpha$, obtained from power law regressions of the data for all sizes and densities, in Fig. \ref{fig2}. First, at large enough area fractions, the exponent $\alpha\to1$, as indicated by the dashed lines in all panels. Second, there is a sublinear regime associated with low-$\varphi$ range, where $\alpha<1$, which is approximately $\varphi\lesssim 0.23$, for all obstacle sizes. In addition, the transition between the sublinear and linear regimes seems to be continuous. Despite the scatter in the data observed in Fig. \ref{fig2}, the linear trend for large $\varphi$ is robust. Finally, we point out that this linear scaling of the vortex mass with the total number of particles implies scaling with the system size as $L^2$, since the area fraction is fixed.

The existence of two distinct scaling regimes suggests that vortex growth results from a competition
between particle capture from the dilute phase and particle escape from the vortex boundary. To
provide a physical interpretation of this crossover, we introduce a minimal kinetic model based on the
balance of incoming and outgoing particle fluxes.

\subsection{A Minimal Kinetic Model for Vortex Growth}
To rationalize the numerical observations, we propose a minimal kinetic description of vortex
growth based on a balance between particle capture from the dilute phase and particle escape from
the vortex boundary. The underlying assumption is that, on time scales much longer than the
microscopic collision time, the vortex behaves as a quasi-stationary condensed aggregate immersed
in a homogeneous dilute background (bulk) of density $\rho_b$.
Throughout this section $\rho$ denotes  a number density (particles per area), while $\varphi$ denotes  the correspoding area fraction, the two being related by $\varphi=(\frac{\pi}{4})\sigma^{2}\rho$, the fluxes are naturally written in  terms of $\rho$, where the comparison with the simulations is made in terms of $\varphi$.
The growth rate of the vortex mass $M(t)$ can then be written as

\begin{equation}
\frac{dM}{dt}=J_{\rm in}-J_{\rm out},
\label{eq_1}
\end{equation}
where $J_{\rm in}$ and $J_{\rm out}$ denote the incoming and outgoing particle fluxes,
respectively.
The incoming flux originates from active particles in the dilute phase that reach the vortex boundary
and become incorporated into the aggregate. Since particles move with self-propulsion speed
$v_0$, the collision rate with the vortex is proportional to the bulk density $\rho_b$, the propulsion
speed, and the effective capture perimeter $\ell_{\rm cap}\simeq2\pi R_{v}$, where $R_{v}$ is the vortex's characteristic radius. We therefore write
\begin{equation}
J_{\rm in}=k_{\rm in}\rho_b v_0 \ell_{\rm cap}=k_{\rm in}\rho_b v_02\pi R_{v},
\label{eq_2}
\end{equation}
where $k_{\rm in}$ is an effective capture coefficient representing the probability that a particle
colliding with the vortex boundary becomes incorporated into the condensed structure (a similar equation is proposed in \cite{mokhtari17}). 

The outgoing flux is associated with particles located at the vortex boundary that escape after their
propulsion direction is reoriented outward by rotational diffusion. Since the characteristic
reorientation rate is controlled by the rotational diffusion coefficient $D_r$, the escape flux is
proportional to the number of particles in the interfacial layer, $M_{\rm surf}=\rho_c2\pi R_{v}\delta$, where $\rho_c$ is the effective density of the vortex and $\delta$ is the interfacial layer thickness, yielding
\begin{equation}
J_{\rm out}=k_{\rm out}D_r M_{\rm surf}=k_{\rm out}D_r\rho_c2\pi R_{v}\delta,
\label{eq_4}
\end{equation}
where $k_{\rm out}$ is an effective escape coefficient that measures the probability that a particle
escapes the aggregate following a rotational reorientation event.

A key feature of eqs. (\ref{eq_2}) and (\ref{eq_4}) is that both fluxes are proportional to the perimeter of the aggregate, so that $R_{v}$ cancel when the stationary condition $\frac{dM}{dt}=0$ is imposed. The balance therefore does not select a preferred vortex size. Instead, it fixes a single value of the background density,
\begin{equation}
\rho_b^{*} = \frac{k_{\rm out}}{k_{\rm in}}
\frac{D_r\delta}{v_0}
\rho_c.
\label{eq_7}
\end{equation}

This quantity 
is the effective bulk density at which particle capture and
particle escape exactly balance each other.

Because the vortex radius has dropped out, $\rho^{*}_{b}$ depends only on the microscopic parameter of the model and not on the size of the aggregate, which is precisely what allows it to act as a threshold: if the dilute background lies below $\rho^{*}_{b}$, losses exceed captures and the aggregate can not sustain extensive growth, whereas above $\rho^{*}_{b}$ the aggregate grows until it depletes the background back to $\rho^{*}_{b}$.

The consequences of this threshold for the system size become explicit when particle conservation is imposed. In the stationary regime,
\begin{equation}
N=M+\rho_b^* (A-A_c),
\label{eq_8}
\end{equation}
where $A$ is total area available to the particles,$A_{c}$ is the area occupied by the condensed phase and  $A_b=A-A_c$ is the area occupied by the dilute phase. Using $M=\rho_{c}A_{c}$ to 
eliminate $A_{c}$, $Ac=\frac{N-\rho^{*}_{b}A}{\rho_{c}-\rho^{*}_{b}}$,  $M=\rho_{c}\frac{N-\rho^{*}_{b}A}{\rho_{c}-\rho^{*}_{b}}$, and introducing the global density $\rho_{t}=\frac{N}{A}$, the vortex mass fraction becomes

\begin{equation}
f_c=\frac{M}{N}=\frac{1-(\rho^{*}_{b}/\rho_{t})}{1-(\rho^{*}_{b}/\rho_{c})}=\frac{1-(\varphi^{*}_{b}/\varphi_{t})}{1-(\varphi^{*}_{b}/\varphi_{c})}.
\label{eq_10}
\end{equation}

The last equality follows because $f_{c}$ is a ratio of densities, where the factor $\frac{\pi}{4}\sigma^{2}$ cancels, so the expression holds unchanged in terms of area fraction, with $\varphi_{t}$ the global area fraction defined in section \ref{Model} and $\varphi_{c}$ that of the condensed phase. When the condensed phase is much denser than the coexisting background, $\varphi_{c}\gg\varphi^{*}_{b}$, eq. (\ref{eq_10}) reduces to
\begin{equation}
f_c
\simeq
1-\frac{\varphi_b^{*}}{\varphi_{t}}.
\label{eq_11}
\end{equation}
Therefore, the model predicts the existence of two distinct regimes. For global densities below the
effective background density $\varphi_b^{*}$, eq. (12) admits no physical solution: the vortex remains  a localized aggregate whose
relative mass vanishes in the infinite size limit in aggregate with the sublinear exponent $\alpha<1$ of Fig. \ref{fig2}. For densities above $\varphi_b^{*}$, 
capture compensates escape  and the obstacle stabilizes an extensive condensed aggregate
containing a finite fraction of all particles in the system, so that $M\propto N$ and $\alpha=1$. In this picture $\varphi_b^{*}$ plays the role
of a kinetic threshold separating localized from extensive vortex states, rather than that of a thermodynamic
coexistence density.

\begin{figure}[b]
\includegraphics[scale=1.0]{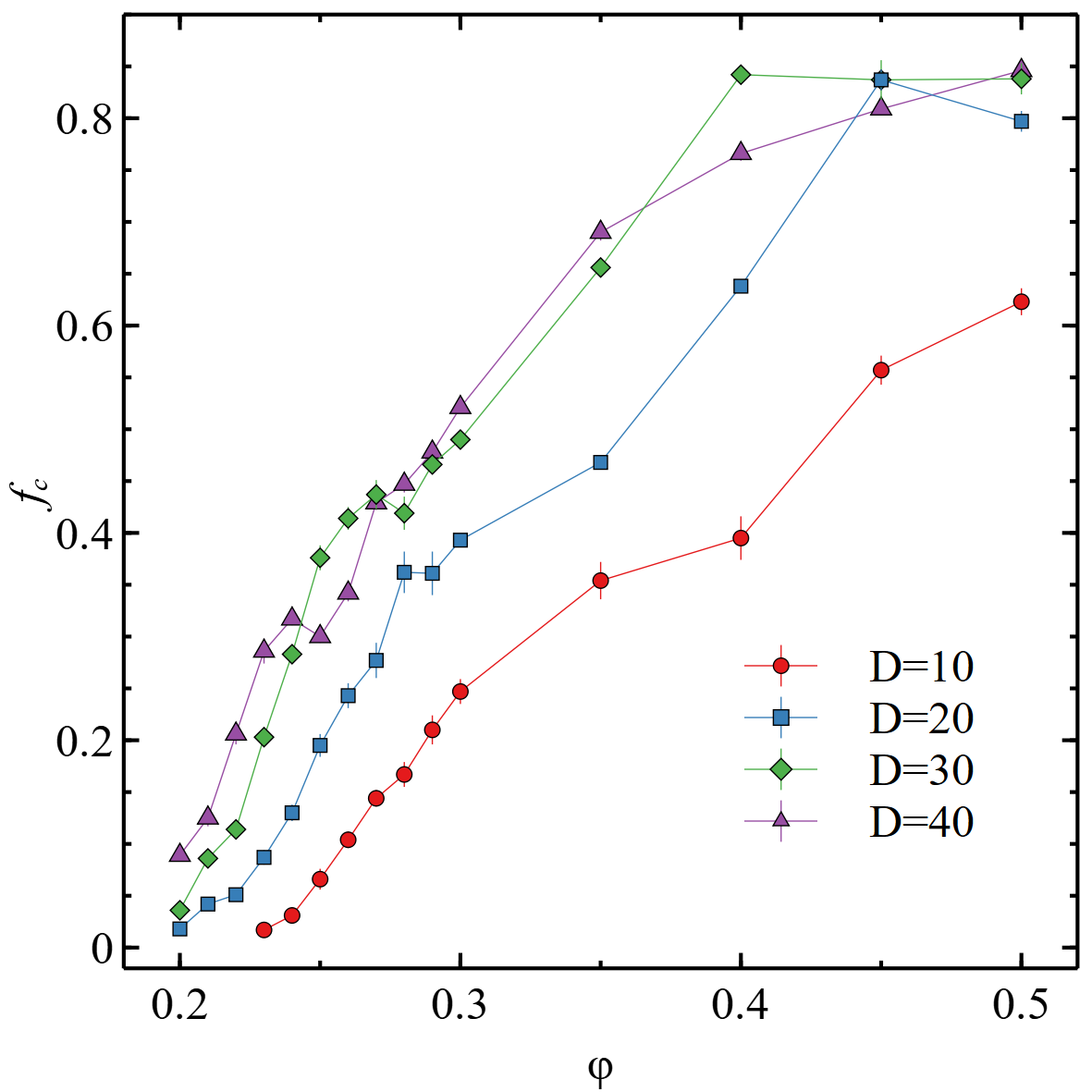}
\caption{Vortex mass fraction, $f_c$, as a function of density, for all obstacle diameters.\label{fig2-1}}
\end{figure}

\begin{figure}[b]
\includegraphics[scale=1.0]{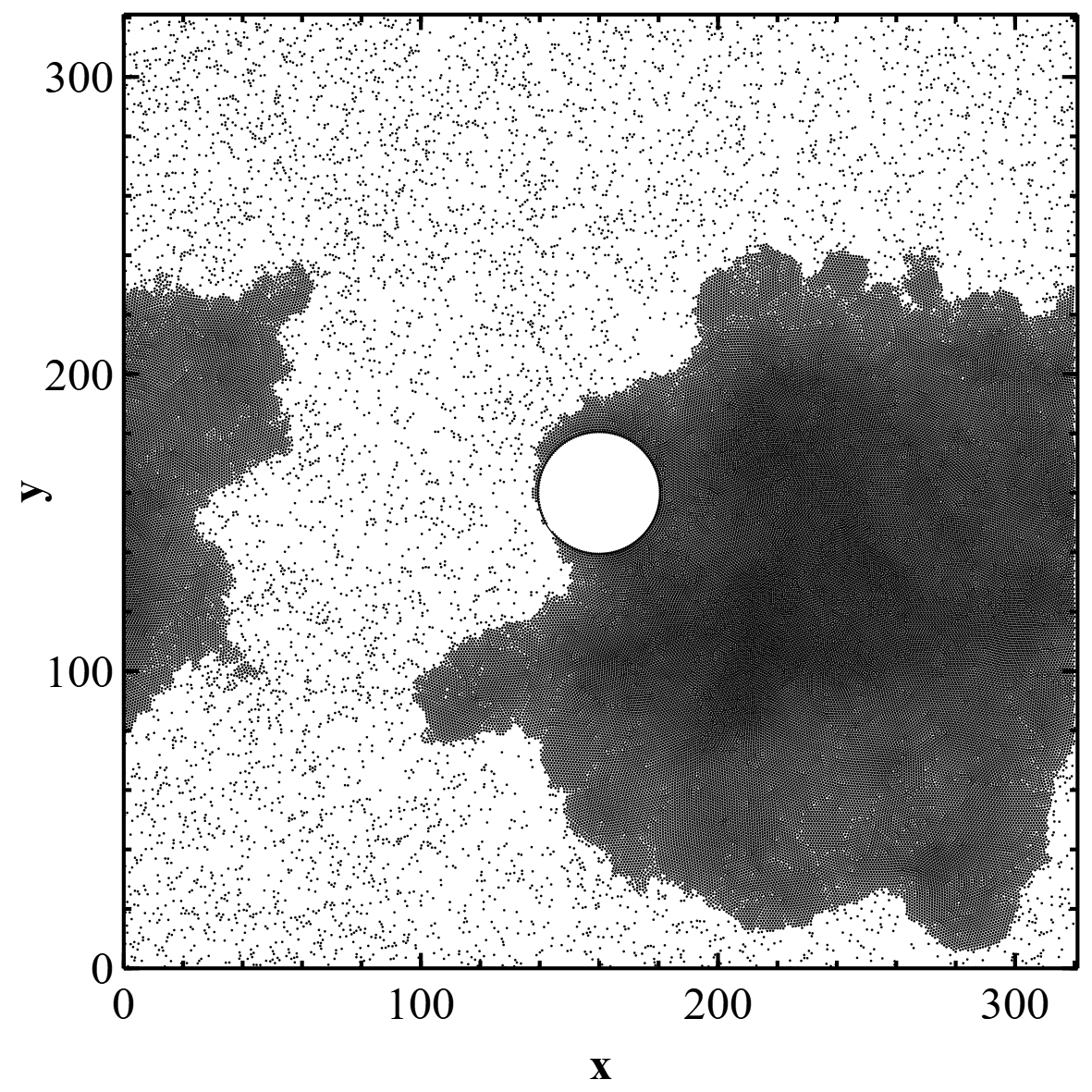}
\caption{System configuration at $\varphi=0.50$, $D=40$, and $L=320$ ($L/D=8$).\label{fig3}}
\end{figure}

We can measure $f_c$ directly from the data for the mean vortex mass assuming a linear relationship between $\overline{M}$ and $N$, and calculating the slope of this curve from linear fits; the results are shown in Fig. \ref{fig2-1}. The data
confirm the trend predicted by eqs. (\ref{eq_10}) and (\ref{eq_11})
that $f_{c}$ increases with  system density,
implying that the vortex incorporates an increasing fraction of the particles, as seen in Fig. \ref{fig1}. Also, we will see that $f_c$ grows from near zero to approximately $0.90$ as we increase $\varphi$.
We emphasize that the present model is not intended as a quantitative theory of vortex growth.
Rather, it provides a minimal kinetic framework that captures the competition between particle
capture and particle escape and offers a physical interpretation for the emergence of localized and
extensive vortex states observed in the simulations.

We finally note that, in very dense and large systems, gas bubbles have been reported  to appear within the condensed phase \cite{xiaqing20}. For the parameters used here, we observed no such bubbles, even for the largest vortices, which  incorporate  nearly all  particles of the system. In any case, $\rho_b^*$ is not expected to vanish in this very dense regime.

\subsection{Cluster-size distributions}
The MIPS is usually identified either by measurements of number fluctuations \cite{fily14,digregorio18} or by measurements of the cluster-size distributions, which develop a peak at large sizes separated from the lower, non-vanishing distribution points. The emergence of a large cluster within the system, which in some instances occurs as percolated stripes  \cite{digregorio18,chacon22}, as irregularly shaped aggregates \cite{caporusso23} or even circular-like structures \cite{fily14,caprini20}, marks the phase separation. In Fig. \ref{fig3}, we show a system configuration for $D=40$, $\varphi=0.50$, and $L/D=8$, within the phase separated regime. Note that the 
vortex has an irregular shape \cite{mokhtari17}. A detailed morphological analysis is beyond the scope of the present work; we only discuss the morphology  briefly below. Besides this irregular shape, it is unevenly distributed around the obstacle surface. Finally, this figure suggests that the phase separation occurs mainly as a result of the vortex, since there is no visible aggregate off the central obstacle. 

\begin{figure}[b]
\includegraphics[scale=1.0]{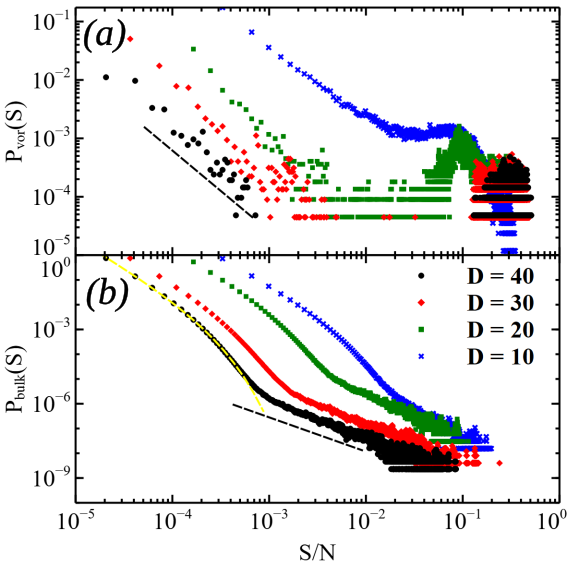}
\caption{Cluster Size Distribution for the vortex, $P_{\mathrm{vor}}(S)$, panel (a), and the bulk, $P_{\mathrm{bulk}}(S)$, panel (b), for different obstacle sizes D, density $\varphi=0.24$ and $L/D=10$, as functions of the normalized cluster size by the total number of particles $S/N$ . The black dashed lines, in both panels, represent the function $y\sim x^{-1.383}$, the yellow dashed line in panel (b) is the function $y\sim x^{-2.35}\exp(-ax)$, with $a=1.55\times10^{-4}$. Both panels are in log-log scale.\label{fig4}}
\end{figure}

The cluster size distribution $P(S)$  gives the mean number of clusters with size (mass) $S$. Here, we measure two distributions: one for the vortex, $P_{\mathrm{vor}}(S)$ [Fig. \ref{fig4}(a)], and another for the bulk, $P_{\mathrm{bulk}}(S)$ [Fig. \ref{fig4}(b)]. We show some examples of both distributions, in log-log plots as functions of the cluster size normalized by the total number of particles, $S/N$, both for $L/D=10$, $\varphi=0.24$, and for different obstacle sizes. 

First, in Fig. \ref{fig4}(a), we see that there is an initial decay of $P_{\mathrm{vor}}(S)$ with $S$, which is approximately independent of the obstacle size. This decay follows a  power law with an exponent, in this case, $P \sim S^{-1.383}$. After this initial regime, the distribution either is approximately constant or zero and eventually exhibits a peak at large $S$. We identify the peaks at large $S$ with the vortices. Similar behavior of $P(S)$ with $S$ was reported in Ref.  \cite{fily14}. 

In typical MIPS scenarios, the cluster size distribution exhibits a clear gap between the small-cluster regime and the peak corresponding to the condensed phase \cite{fily14}. In our system, this gap corresponds to the initial power-law decay and the large vortex cluster, and it only becomes apparent for sufficiently large obstacle sizes ($D\geq 30$) even though the system is already in the phase separated regime based on the scaling exponent of $\overline{M}$. In contrast, for $D\leq20$, the gap is not clearly visible, which indicates a transitional regime in which  vortex growth is present but the morphological separation between small clusters and the vortex is not fully established. Hence, we conclude that the emergence of a large vortex, particularly for $D\leq 20$, is not clearly distinguished from smaller clusters that may form near the obstacle. In other words, the vortex mass exhibits strong fluctuations, which hinders the clear identification of a dominant large cluster.  
Finally, we note that, at this area fraction, the vortex contains only about $30\%$ of the particles in the system, which is significantly lower than the values reported in Ref. \cite{fily14}, where the clusters in the MIPS incorporate nearly all particles. This point is discussed in more detail below.

 The initial decay of the bulk cluster size distributions  for small $S$ is well fitted by a function of the form $y\sim\mathrm{e}^{-ax}/x^b$ (see the yellow dashed line in Fig. \ref{fig4}(b)), with the constants given by $a=1.55\times10^{-4}$ and $b=2.35$. This exponential form was also previously reported, but with $b=1$ \cite{fily14}. More complex fitting functions, such as the Weibull distribution \cite{caporusso23} were also reported for the size distribution, with the fitting parameters being related to the cluster's radius of gyration. The bulk size distributions $P_{\mathrm{bulk}}(S)$ change their behavior at intermediate cluster sizes, where they decay more slowly, following a power law with an exponent very close to that observed for $P_{\mathrm{vor}}(S)$ at small sizes. Although the origin of this similarity remains unclear and it may be coincidental, we observe that the decay exponents of $P_{\mathrm{bulk}}(S)$ at intermediate sizes and $P_{\mathrm{vor}}(S)$ at small sizes are very similar for all $D$, $\varphi$ and large $L/D$. Despite the  presence of a finite, albeit small, number of large clusters in the bulk, for instance, for $D=10$, there is a finite probability of observing a cluster containing about $20\%$ of $N$; however, there is no peak clearly separated from the rest of the distribution. Hence, the phase separated state is solely determined by the vortex when the global area fraction is less than the known MIPS critical value. This supports the interpretation that the vortex constitutes the condensed cluster in the phase-separated states.

\begin{figure}[b]
\includegraphics[scale=1.0]{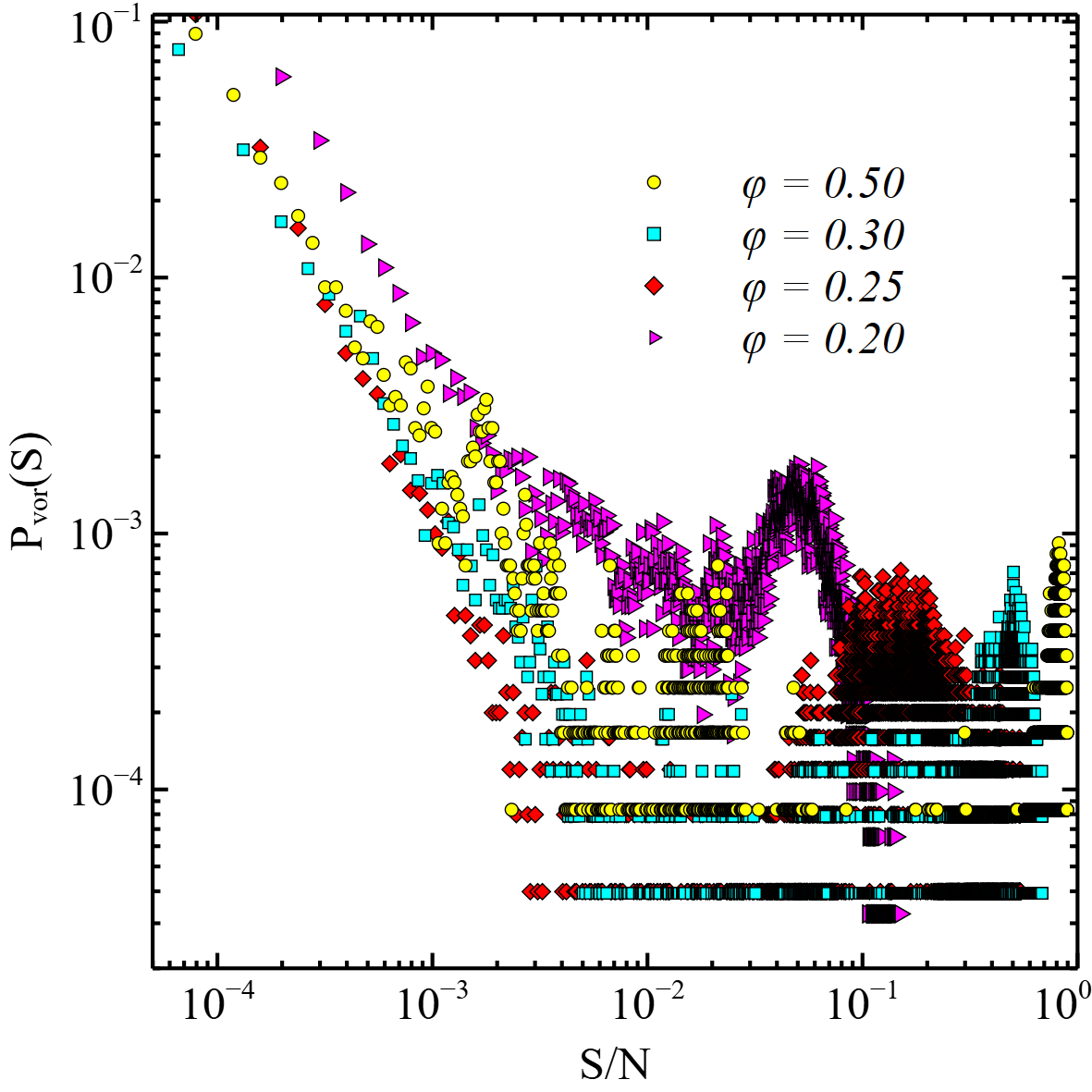}
\caption{Vortex Size Distribution, $P_{\mathrm{vor}}(S)$, for $D=20$, $L/D=10$, and different densities as functions of the  cluster size normalized by the total number of particles $S/N$. \label{fig5}}
\end{figure}

\begin{figure}[b]
\includegraphics[scale=1.0]{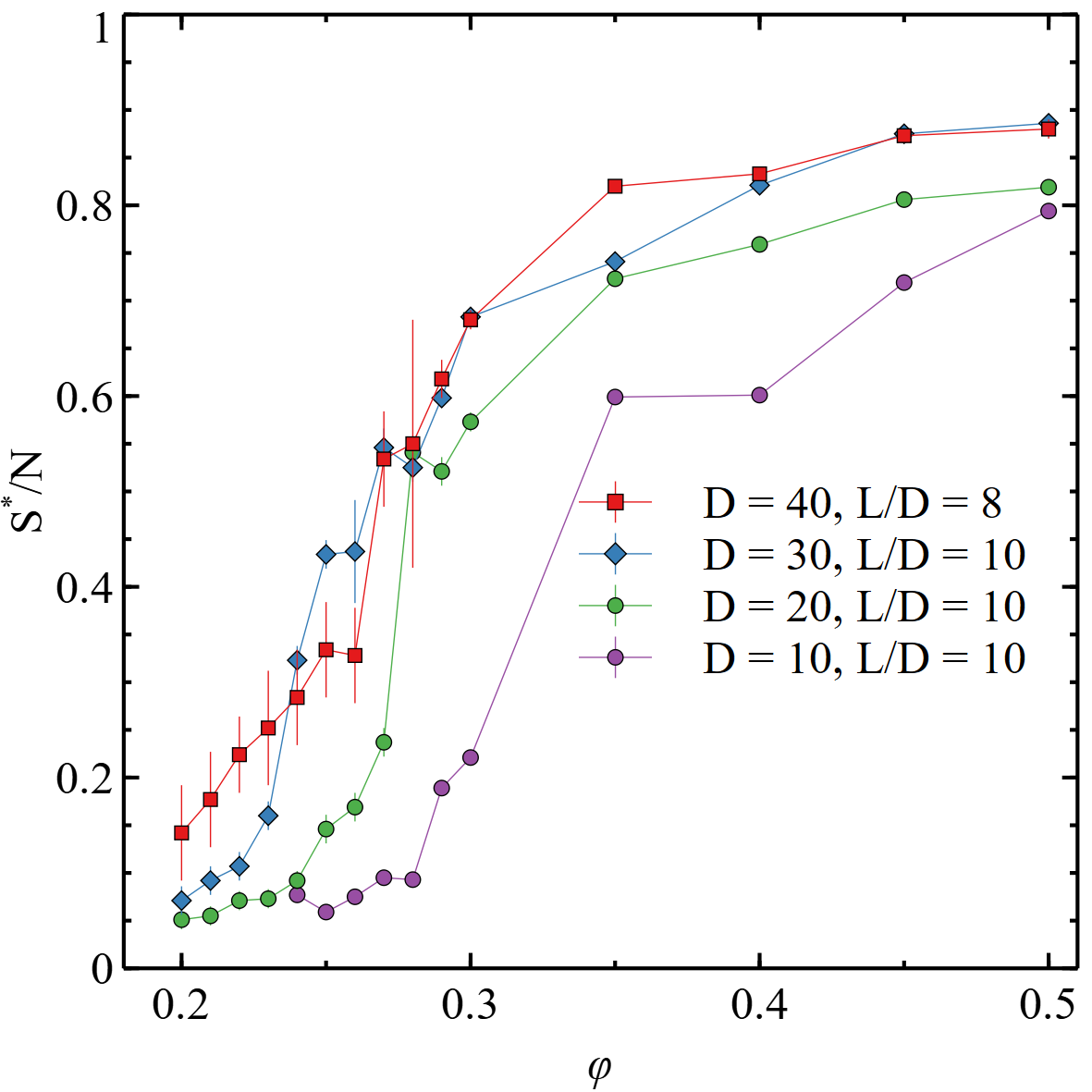}
\caption{Cluster size at the peak of $P_{\mathrm{vor}}(S)$, $S^*/N$, for $D=10$, $20$ and $30$, $L/D=10$, and $D=40$, $L/D=8$, as a function of $\varphi$. \label{fig6}}
\end{figure}

Regarding the behavior of the vortex size with density, as aforementioned, the vortex does not contain most of the particles in the system at the onset of the phase separated state. Actually, the vortex mass fraction  increases with $D$, as seen in Fig. \ref{fig5}, which shows the vortex size distribution for $D=20$, $L/D=10$ and different $\varphi$. We see that the overall shape of the distribution does not change with $\varphi$.  However, as $\varphi$ increases, the gap between the two regimes becomes larger and, as a consequence, the peak position $S^*$, which characterizes  the vortex size, shifts  toward large $S/N$. We show in Fig. \ref{fig6} the peak position $S^*$ of the vortex size distribution, as a function of $\varphi$ for all obstacle sizes and largest $L/D$ for each $D$. We see that the vortex size grows significantly at specific area fractions for distinct $D$, but all of these values are larger than $0.23$, which is our estimated density for the obstacle-induced phase separated regime; for instance, for $D=20$, this growth is only seen for $\varphi>0.27$, while at $D=10$, the large growth occurs at $\varphi=0.30$. Usually, after this fast growth of $S^*$, the vortex contains approximately $50\%$ of the system particles. This fact confirms the information shown in Fig. \ref{fig4}(b), namely, that there is no giant cluster in the bulk, since most of the particles belong to the vortex. At much larger $L/D$ ratios, large clusters may perhaps appear in the bulk, separated from the vortex. However, if the current trend persists up to this regime, such clusters are expected to remain small relative to the vortex, so  that MIPS would be mainly associated with the latter aggregate.


\subsection{Mean and Maximum Vortex Radii}
Here, we present and discuss our results for both the mean vortex radius, Eq. (\ref{eqn4}), and the maximum vortex radius, Eq. (\ref{eqn5}).

\begin{figure}[b]
\includegraphics[scale=1.0]{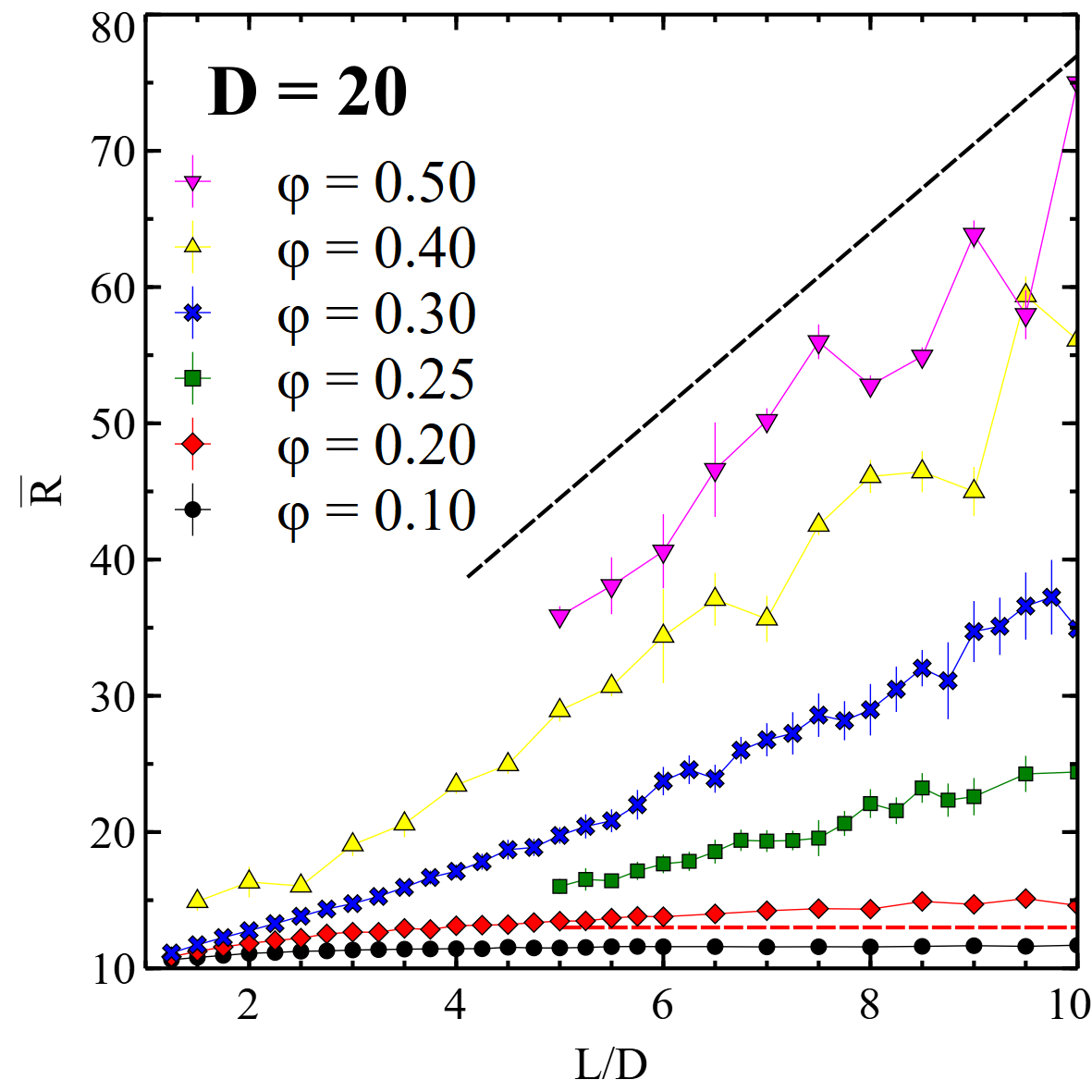}
\caption{Mean vortex radius, $\overline{R}$ as a function of the ratio $L/D$, the ratio of the system length to the obstacle diameter, for $D=20$ and some density values. The black and red dashed lines are $\overline{R}\sim L/D$ and $\overline{R}=\mathrm{const}$, respectively. \label{fig7}}
\end{figure}

In Fig. \ref{fig7}, we show the mean vortex radius, $\overline{R}$, for $D=20$ as a function of the ratio between the system size and the obstacle diameter, $L/D$. Similarly to the results for the vortex mass (see Fig. \ref{fig1}), two distinct regimes emerge in the dependence between these variables: at low $\varphi$, the mean radius saturates with $L/D$; at higher $\varphi$, it approaches the linear dependence on $L/D$. Analogously to the vortex mass behavior, the transition between regimes seems to be continuous. We observed similar results for the other obstacle diameters.

\begin{figure}[b]
\includegraphics[scale=1.0]{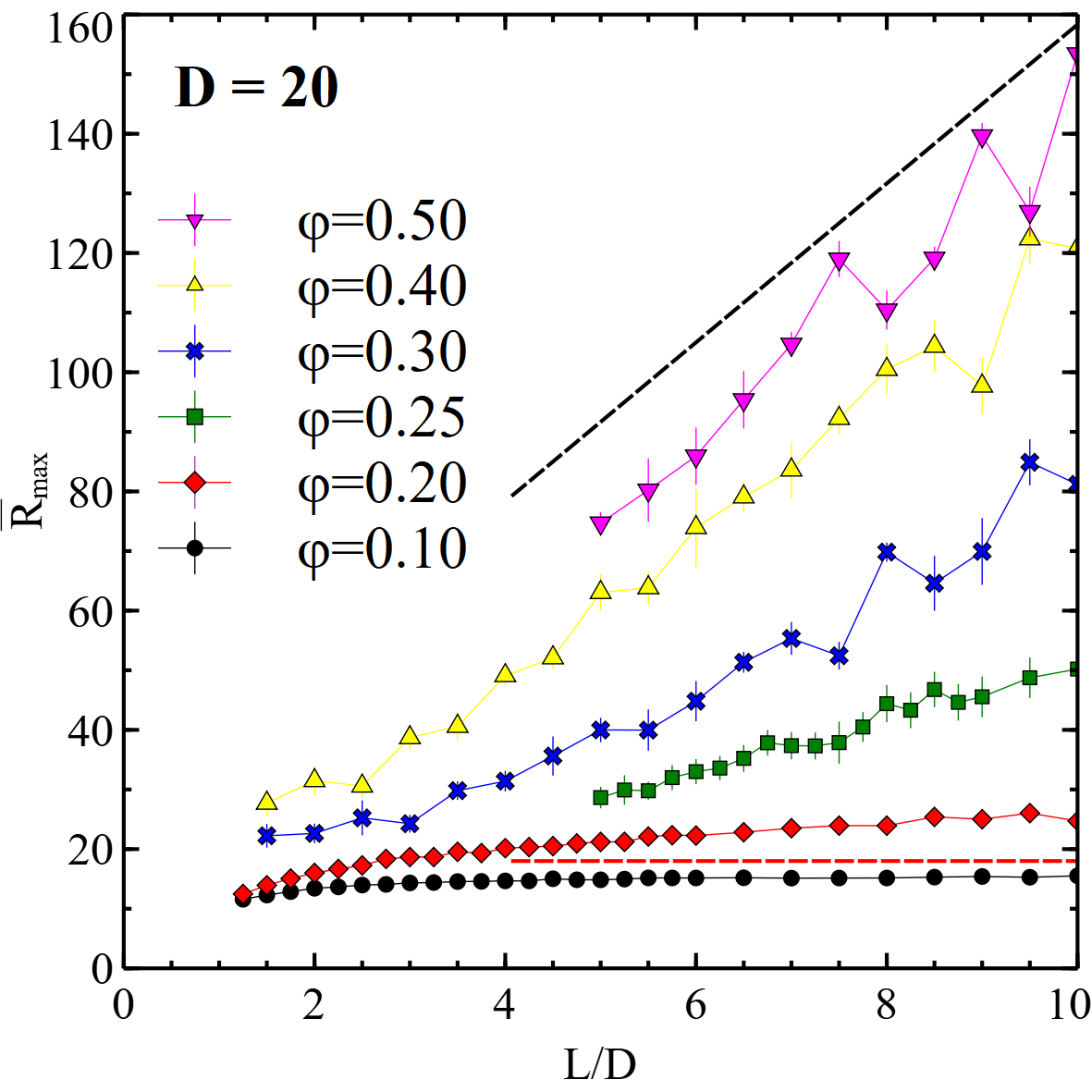}
\caption{Maximum vortex radius, $\overline{R}_{\mathrm{max}}$ as a function of $L/D$ for $D=20$ and some density values. The black and red dashed lines are $\overline{R}_{max}\sim L/D$ and $\overline{R}_{max}=\mathrm{const}$, respectively.\label{fig8}}
\end{figure}

The results for the maximum vortex radius, $\overline{R}_{\mathrm{max}}$, as a function of $L/D$ are shown in Fig. \ref{fig8}, again for $D=20$. The same trends observed for the vortex mass and mean radius are present here: at low $\varphi$, the maximum radius saturates, while at high $\varphi$, it increases linearly with $L$. 


We can speculate about the morphology of the vortex. If a vortex has a circular distribution of particles, which has a minimum radius of $D/2$ and a maximum radius $\overline{R}_{max}$, we have $\overline{R}=(\overline{R}_{max}+D/2)/2$; for an irregularly shaped vortex, like the one seen in Fig. \ref{fig3}, the relationship between these two mean distances depends on the shape. Our data indicate the ratio $\overline{R}_{max}/\overline{R}\approx2$ for the largest $L/D$ values we studied; in this limit, we can safely neglect the $D/2$ term. On the other hand, we also measured the ratio between the two eigenvalues of the inertia tensor $g_{xx}$ and $g_{yy}$, and it is significantly distinct from $1$. Hence, we conclude that, even though the  vortex shape is generally noncircular, its mass is approximately evenly distributed along the radial distance from the obstacle,  closer to a rotating slab than to a rotating cog.


\section{Conclusions\label{Conclusions}}
We studied the properties of a vortex of dry active matter formed around a circular obstacle. We measured its mean total mass, mean maximum radius, and mean radius as functions of the system size $L$, obstacle diameter $D$, and area  fraction $\varphi$. Our main goal was to determine  how these quantities scale with the system size and to understand how the spontaneous phase-separation scenario is modified by the presence of the obstacle. We focused our conclusions mainly on the results in the large system size limit, $L/D\gg1$.

Our results  indicated that the kinetically stabilized condensate occurs for $D\geq 30$ for a sufficiently large  obstacle, since the exponent of the power law relation between the mean mass and the total number of particles was approximately unity in this regime. Below this area fraction, the vortex mass grew more slowly with $N$,  indicating  that no stable vortex persisted in the infinite system size limit, $L\to\infty$. Moreover, from the analysis of the cluster size distributions  for the vortex, we observed that the large cluster  identified as the vortex was not clearly separated from smaller aggregates for $D\leq20$ at any value of $\varphi$. For larger $\varphi$, we saw that such a sharp distinction between small and large aggregates in the vortex became clear only for $\varphi\geq0.30$. Hence, phase separation was clearly visible only for sufficiently large obstacles. The size distributions also indicated that the emerging vortex  contained only a  small fraction  of the total number of particles. As we increased the area fraction, this amount increased, reaching nearly  $93\%$ of $N$ for $\varphi=0.50$ and $D=20$. 

Regarding the mean and maximum radii we observed that they also followed trends with the system size that were very similar to those found for the mean mass as a function of the total number of particles. Finally, the analysis of  the ratio of the mean and maximum radii and the two eigenvalues of the inertial tensor revealed that the vortex most likely exhibited a slab-like geometry.
In summary, our results showed that obstacle-induced phase separation occurred only above minimum values of obstacle size and density. Below these thresholds, vortex formation was incomplete or transitional; at higher densities and for larger obstacles, the vortex became the main condensed region, although it retained an irregular shape. Those results may assist future studies and demonstrate how geometric confinement can be used to control aggregation in active matter systems.

\section{Acknowledgments}
This study was financed in part by the Coordenação de Aperfeiçoamento de Pessoal de Nível Superior-Brasil (CAPES) -Finance Code 001. J. L. C. D. acknowledges funding from the European Union’s Horizon Europe ERA Fellowship project (PattSpin, Grant No. 101130777).
W. P. Ferreira acknowledges funding from Conselho Nacional de Desenvolvimento Científico e Tecnológico - Brasil (CNPq) -Grant No. 309252/2025-3.


\begin{thebibliography}{99}
 \bibitem{marchetti13}
M. C. Marchetti, J. F. Joanny, S. Ramaswamy, T. B. Liverpool, J. Prost, M. Rao, and R. A. Simha,
Hydrodynamics of soft active matter,
\href{https://doi.org/10.1103/RevModPhys.85.1143}{Rev. Mod. Phys. 85, 1143 (2013)}.

\bibitem{tevrugt24}
M. te Vrugt and R. Wittkowski,
Metareview: a survey of active matter reviews,
\href{https://doi.org/10.1140/epje/s10189-024-00466-z}{Eur. Phys. J. E 48, 12 (2024)}.

\bibitem{volpe25}
G. Volpe, N. A. Araújo, M. Guix, M. Miodownik, N. Martin, L. Alvarez, J. Simmchen, et al.,
Roadmap for animate matter,
\href{https://iopscience.iop.org/article/10.1088/1361-648X/adebd3/meta}{J. Phys.: Condens. Matter 37, 333501 (2025)}.

\bibitem{berke08}
A. P. Berke, L. Turner, H. C. Berg, and E. Lauga,
Hydrodynamic attraction of swimming microorganisms by surfaces,
\href{https://link.aps.org/doi/10.1103/PhysRevLett.101.038102}{Phys. Rev. Lett. 101, 038102 (2008)}.


\bibitem{fily12}
Y. Fily and M. C. Marchetti,
Athermal phase separation of self-propelled particles with no alignment,
\href{https://link.aps.org/doi/10.1103/PhysRevLett.108.235702}{Phys. Rev. Lett. 108, 235702 (2012)}.

\bibitem{redner13}
G. S. Redner, M. F. Hagan, and A. Baskaran,
Structure and dynamics of a phase-separating active colloidal fluid,
\href{https://link.aps.org/doi/10.1103/PhysRevLett.110.055701}{Phys. Rev. Lett. 110, 055701 (2013)}.

\bibitem{buttinoni13}
I. Buttinoni, J. Bialké, F. Kümmel, H. Löwen, C. Bechinger, and T. Speck,
Dynamical clustering and phase separation in suspensions of self-propelled colloidal particles,
\href{https://link.aps.org/doi/10.1103/PhysRevLett.110.238301}{Phys. Rev. Lett. 110, 238301 (2013)}.

\bibitem{fily14}
Y. Fily, S. Henkes, and M. C. Marchetti,
Freezing and phase separation of self-propelled disks,
\href{http://dx.doi.org/10.1039/C3SM52469H}{Soft Matter 10, 2132 (2014)}.

\bibitem{cates15}
M. E. Cates and J. Tailleur,
Motility-induced phase separation,
\href{https://www.annualreviews.org/content/journals/10.1146/annurev-conmatphys-031214-014710}{Annu. Rev. Condens. Matter Phys. 6, 219 (2015)}.

\bibitem{stenhammar15}
J. Stenhammar, R. Wittkowski, D. Marenduzzo, and M. E. Cates,
Activity-induced phase separation and self-assembly in mixtures of active and passive particles,
\href{https://doi.org/10.1103/PhysRevLett.114.018301}{Phys. Rev. Lett. 114, 018301 (2015)}.

\bibitem{wittmann16}
R. Wittmann and J. M. Brader,
Active Brownian particles at interfaces: An effective equilibrium approach,
\href{https://iopscience.iop.org/article/10.1209/0295-5075/114/68004/meta}{Europhys. Lett. 114, 68004 (2016)}.

\bibitem{liebchen17}
B. Liebchen and D. Levis,
Collective behavior of chiral active matter: Pattern formation and enhanced flocking,
\href{https://link.aps.org/doi/10.1103/PhysRevLett.119.058002}{Phys. Rev. Lett. 119, 058002 (2017)}.

\bibitem{digregorio18}
P. Digregorio, D. Levis, A. Suma, L. F. Cugliandolo, G. Gonnella, and I. Pagonabarraga,
Full phase diagram of active Brownian disks: From melting to motility-induced phase separation,
\href{https://link.aps.org/doi/10.1103/PhysRevLett.121.098003}{Phys. Rev. Lett. 121, 098003 (2018)}.

\bibitem{caprini20}
L. Caprini, U. Marini Bettolo Marconi, and A. Puglisi,
Spontaneous velocity alignment in motility-induced phase separation,
\href{https://link.aps.org/doi/10.1103/PhysRevLett.124.078001}{Phys. Rev. Lett. 124, 078001 (2020)}.

\bibitem{xiaqing20}
X.-q. Shi, G. Fausti, H. Chaté, C. Nardini, and A. Solon,
Self-organized critical coexistence phase in repulsive active particles,
\href{https://link.aps.org/doi/10.1103/PhysRevLett.125.168001}{Phys. Rev. Lett. 125, 168001 (2020)}.

\bibitem{su2021}
J. Su, H. Jiang, and Z. Hou,
Inertia-induced nucleation-like motility-induced phase separation,
\href{https://dx.doi.org/10.1088/1367-2630/abd80a}{New J. Phys. 23, 013005 (2021)}.

\bibitem{chacon22}
E. Chacón, F. Alarcón, J. Ramírez, P. Tarazona, and C. Valeriani,
Intrinsic structure perspective for MIPS interfaces in two-dimensional systems of active Brownian particles,
\href{http://dx.doi.org/10.1039/D1SM01493E}{Soft Matter 18, 2646 (2022)}.

\bibitem{rojasvega23}
M. Rojas-Vega, P. de Castro, and R. Soto,
Wetting dynamics by mixtures of fast and slow self-propelled particles,
\href{https://link.aps.org/doi/10.1103/PhysRevE.107.014608}{Phys. Rev. E 107, 014608 (2023)}.

\bibitem{caporusso23}
C. B. Caporusso, L. F. Cugliandolo, P. Digregorio, G. Gonnella, D. Levis, and A. Suma,
Dynamics of motility-induced clusters: Coarsening beyond Ostwald ripening,
\href{https://link.aps.org/doi/10.1103/PhysRevLett.131.068201}{Phys. Rev. Lett. 131, 068201 (2023)}.


\bibitem{pan20}
J.-x. Pan, H. Wei, M.-j. Qi, H.-f. Wang, J.-j. Zhang, W.-d. Tian, and K. Chen,
Vortex formation of spherical self-propelled particles around a circular obstacle,
\href{http://dx.doi.org/10.1039/D0SM00277A}{Soft Matter 16, 5545 (2020)}.

\bibitem{bendor22}
Y. Ben Dor, S. Ro, Y. Kafri, M. Kardar, and J. Tailleur,
Disordered boundaries destroy bulk phase separation in scalar active matter,
\href{https://link.aps.org/doi/10.1103/PhysRevE.105.044603}{Phys. Rev. E 105, 044603 (2022)}.



\bibitem{sepulveda17}
N. Sepúlveda and R. Soto,
Wetting transitions displayed by persistent active particles,
\href{https://link.aps.org/doi/10.1103/PhysRevLett.119.078001}{Phys. Rev. Lett. 119, 078001 (2017)}.

\bibitem{sepulveda18}
N. Sepúlveda and R. Soto,
Universality of active wetting transitions,
\href{https://link.aps.org/doi/10.1103/PhysRevE.98.052141}{Phys. Rev. E 98, 052141 (2018)}.


\bibitem{turci21}
F. Turci and N. B. Wilding,
Wetting transition of active Brownian particles on a thin membrane,
\href{https://link.aps.org/doi/10.1103/PhysRevLett.127.238002}{Phys. Rev. Lett. 127, 238002 (2021)}.

\bibitem{bechinger16}
C. Bechinger, R. Di Leonardo, H. Löwen, C. Reichhardt, G. Volpe, and G. Volpe,
Active particles in complex and crowded environments,
\href{https://link.aps.org/doi/10.1103/RevModPhys.88.045006}{Rev. Mod. Phys. 88, 045006 (2016)}.

\bibitem{makarchuk19}
S. Makarchuk, V. C. Braz, N. A. M. Araújo, L. Ciric, and G. Volpe,
Enhanced propagation of motile bacteria on surfaces due to forward scattering,
\href{https://doi.org/10.1038/s41467-019-12010-1}{Nature Communications 10, 4110 (2019)}.

\bibitem{doostmo16}
A. Doostmohammadi, M. F. Adamer, S. P. Thampi, and J. M. Yeomans,
Stabilization of active matter by flow-vortex lattices and defect ordering,
\href{https://doi.org/10.1038/ncomms10557}{Nature Communications 7, 10557 (2016)}.


\bibitem{galajda07}
P. Galajda, J. Keymer, P. Chaikin, and R. Austin,
A wall of funnels concentrates swimming bacteria,
\href{https://journals.asm.org/doi/abs/10.1128/jb.01033-07}{Journal of Bacteriology 189, 8704 (2007)}.

\bibitem{mahmud09}
G. Mahmud, C. J. Campbell, K. J. M. Bishop, Y. A. Komarova, O. Chaga, S. Soh, S. Huda, K. Kandere-Grzybowska, and B. A. Grzybowski,
Directing cell motions on micropatterned ratchets,
\href{https://doi.org/10.1038/nphys1306}{Nature Physics 5, 606 (2009)}.

\bibitem{wioland16}
H. Wioland, F. G. Woodhouse, J. Dunkel, and R. E. Goldstein,
Ferromagnetic and antiferromagnetic order in bacterial vortex lattices,
\href{https://doi.org/10.1038/nphys3607}{Nature Physics 12, 341 (2016)}.

\bibitem{heidenreich16}
S. Heidenreich, J. Dunkel, S. H. L. Klapp, and M. Bär,
Hydrodynamic length-scale selection in microswimmer suspensions,
\href{https://link.aps.org/doi/10.1103/PhysRevE.94.020601}{Phys. Rev. E 94, 020601 (2016)}.

\bibitem{james21}
M. James, D. A. Suchla, J. Dunkel, and M. Wilczek,
Emergence and melting of active vortex crystals,
\href{https://doi.org/10.1038/s41467-021-25545-z}{Nature Communications 12, 5630 (2021)}.

\bibitem{bricard15}
A. Bricard, J.-B. Caussin, D. Das, C. Savoie, V. Chikkadi, K. Shitara,
O. Chepizhko, F. Peruani, D. Saintillan, and D. Bartolo,
Emergent vortices in populations of colloidal rollers,
\href{https://doi.org/10.1038/ncomms8470}{Nature Communications 6, 7470 (2015)}.


\bibitem{schimming25}
C. D. Schimming, C. J. O. Reichhardt, and C. Reichhardt,
Vortex lattices in active nematics with periodic obstacle arrays,
\href{https://link.aps.org/doi/10.1103/PhysRevLett.132.018301}{Phys. Rev. Lett. 132, 018301 (2024)}.

\bibitem{tiwari24}
C. Tiwari and S. P. Singh,
Collective dynamics of active dumbbells near a circular obstacle,
\href{https://doi.org/10.1039/d4sm00044g}{Soft Matter 20, 4816 (2024)}.

\bibitem{potiguar14}
F. Q. Potiguar, G. A. Farias, and W. P. Ferreira,
Self-propelled particle transport in regular arrays of rigid asymmetric obstacles,
\href{https://link.aps.org/doi/10.1103/PhysRevE.90.012307}{Phys. Rev. E 90, 012307 (2014)}.


\bibitem{mokhtari17}
Z. Mokhtari, T. Aspelmeier, and A. Zippelius,
Collective rotations of active particles interacting with obstacles,
\href{https://dx.doi.org/10.1209/0295-5075/120/14001}{Europhys. Lett. 120, 14001 (2017)}.

\bibitem{junior24}
F. P. Júnior, J. L. Domingos, W. Ferreira, and F. Potiguar,
Correlations between two vortices in dry active matter,
\href{https://www.sciencedirect.com/science/article/pii/S0378437124006903}{Physica A 656, 130181 (2024)}.

\bibitem{junior26}
F. P. Júnior, J. L. Domingos, W. Ferreira, and F. Potiguar,
Controlling vortex rotation with obstacles in nonaligning dry active matter,
\href{https://doi.org/10.1103/sjy3-qzp8}{Phys. Rev. E 113, 054119 (2026)}.

\bibitem{honeycutt92}
R. L. Honeycutt,
Stochastic Runge-Kutta algorithms. I. White noise,
\href{https://link.aps.org/doi/10.1103/PhysRevA.45.600}{Phys. Rev. A 45, 600 (1992)}.



 \end{thebibliography}

\end{document}